\documentclass[aps,pra,reprint,amsmath,amssymb,longbibliography,floatfix]{revtex4-2}
\usepackage[T1]{fontenc}
\usepackage[utf8]{inputenc}
\usepackage[english]{babel}
\usepackage{lmodern}
\usepackage{graphicx,bm,booktabs}
\usepackage{xcolor}

\usepackage[colorlinks=true,linkcolor=blue,citecolor=blue,urlcolor=blue]{hyperref}
\newcommand{\bs}{\boldsymbol{s}}
\newcommand{\br}{\boldsymbol{r}}
\newcommand{\bx}{\boldsymbol{x}}
\newcommand{\bv}{\boldsymbol{v}}
\newcommand{\iu}{\mathrm{i}}
\newcommand{\dd}{\mathrm{d}}
\begin{document}
\title{Complex chaotic Boltzmann machines as deterministic representations of quantum states}
\author{J. Montes}
\email{javier.montesm@upm.es}
\affiliation{Grupo de Sistemas Complejos, ETSIME, Universidad Politécnica de Madrid, Rios Rosas 21, 28003 Madrid, Spain}
\begin{abstract}
Neural networks and tensor networks have opened new ways to explore quantum systems by parameterizing families of wave functions through the structure and parameters of a network. Here we extend this perspective to a dynamical representation: we introduce a complex chaotic Boltzmann machine that assigns a deterministic system to a pure quantum state. We extend Suzuki's construction to incorporate both Born probabilities and relative phases. The probabilities set the trajectory velocities in a hypercube, while the phases define a complex field that lifts its Poincar\'e sections. The evolution is determined by the network parameters and initial conditions, without random choices during flights or configuration changes. Our analysis of two qubits and study of three-qubit families show how relations between probabilities organize regular curves, islands, and unstable regions, and how phases distinguish states that share the same Born dynamics. This architecture turns a wave-function parameterization into a geometry of trajectories, providing a tool to explore probabilities, phases, and symmetries together across families of states.
\end{abstract}
\maketitle
\raggedbottom

\begin{figure*}[t]
\centering\includegraphics[width=\textwidth]{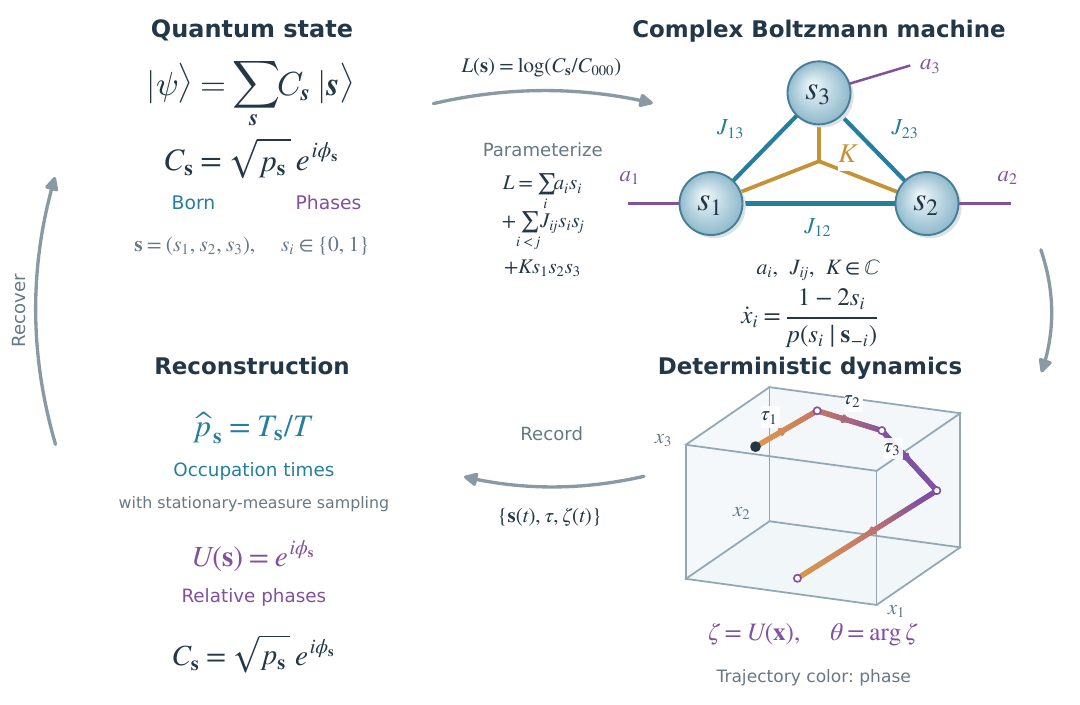}
\caption{CCBM architecture for three qubits. The state determines the complex biases $a_i$, couplings $J_{ij}$, and the gold three-body connection $K$. Suzuki's dynamics ($c_i=1$) generates the flights and impacts; color represents $\theta=\arg U$. The four flights illustrate this process. To reconstruct Born probabilities, $T_{\bs}$ accumulates occupation times and $T$ is the total time; sampling the stationary measure is required. The complex field preserves relative phases at its vertices, which can also be recovered from sufficient records. Together, both parts determine the state up to a global phase.}
\label{fig:architecture}
\end{figure*}

\section{Introduction}\label{sec:intro}
Representing many-body wave functions requires organizing a number of coefficients that grows exponentially with system size. Tensor networks and neural quantum states (NQS) have shown that structured families can be studied through parameterizations adapted to their correlations and symmetries~\cite{Schollwock2011,Carleo2017,Medvidovic2024}. Their parameters span entire families of states and, when the structure allows it, provide a more compact description than the full list of amplitudes. This perspective has expanded the tools available for studying quantum systems and for relating a representation's architecture to the properties of the states it describes~\cite{Deng2017,Chen2018,Glasser2018,Choo2018}.

The possibility of representing a state through a network motivates this work. We propose an architecture in which the complex parameters of a Boltzmann machine specify both a wave function and a deterministic dynamical system. A family of states can then be explored by varying the machine parameters and following changes in its trajectories and Poincar\'e sections. The methodological value lies in a geometry where relations between probabilities, phases, and symmetries appear as structures of the motion.

Our starting point is Suzuki's chaotic Boltzmann machine~\cite{Suzuki2013CBM,Suzuki2013Billiards}. In this construction, the conditional probabilities of a distribution set the velocities of a point moving inside a hypercube. Motion is linear between impacts, and the velocities change when a face is reached. Binary variables indicate the direction of motion of each coordinate. Once initial conditions are chosen, flights and configuration changes proceed without random choices. This rule turns a probability distribution into a system whose trajectories may be regular or chaotic. For two elements, Suzuki reduced the motion to a torus and established conditions for sampling the distribution~\cite{Suzuki2014Two}.

We extend this construction to represent pure quantum states. The complex chaotic Boltzmann machine (CCBM) retains Suzuki's velocity rule for Born probabilities and incorporates a complex field built from the relative phases. Its argument provides the angular coordinate of the lifted sections. The network's complex biases and couplings thus determine both parts of the wave function and its dynamical representation. The construction is defined directly through these parameters and also accepts coefficients obtained from other representations, including a complex restricted Boltzmann machine (RBM).

Figure~\ref{fig:architecture} illustrates the construction. We start from a wave function and use its coefficients to determine the complex biases and couplings of the network. These parameters define the motion inside the cube and the phase field along each trajectory. The diagram closes with the recovery of the original state: occupation times give the Born probabilities when the stationary measure is sampled, and the complex field provides the relative phases. Together, probabilities and phases reconstruct the wave function.

Existing hardware implementations of chaotic Boltzmann machines provide a further motivation for this extension. Yamaguchi and collaborators, including Suzuki, presented a complementary metal-oxide-semiconductor (CMOS) circuit for CBMs and three-neuron network operation~\cite{Yamaguchi2017CMOS}; Kawashima and collaborators implemented CBMs on field-programmable gate arrays (FPGAs) with algorithms adapted to the hardware~\cite{Kawashima2020FPGA}. These precedents suggest a possible implementation combining a CBM circuit for Born dynamics with a module that evaluates the complex phase field from positions and state phases. Together, their records could be used to reconstruct the state and calculate probabilities in other bases, connecting the proposed representation to dedicated platforms for classical emulation of quantum-state families. Such an implementation would require the appropriate conditional probabilities, including higher-order interactions when needed, and an assessment of the cost and accuracy of the phase calculation. Hardware realization of the CCBM and any computational advantage remain to be established.

The trajectory representation complements other ways to organize quantum information, including quasiprobability distributions~\cite{Wigner1932,Ferrie2011} and geometric formulations of states and their evolution~\cite{Heslot1985,Ashtekar1997}. Here the motion explores the representation of a fixed wave function. Its time is auxiliary and does not correspond to Schr\"odinger evolution or to a law of motion for quantum particles such as Bohm's theory~\cite{Bohm1952}. The construction is also related to deterministic sampling~\cite{Chen2016,Neal2012} and samplers for neural quantum states~\cite{Sharir2020,HibatAllah2020,Bagrov2021,Czischek2019}. The absence of random choices during evolution allows each flight and transition to be tracked; recovering probabilities from occupation times requires the sampling conditions specified below.

To use Suzuki's machine as a representation of a quantum state, we must include relative phases as well as Born probabilities. We therefore extend it to a complex chaotic Boltzmann machine, define how these phases enter its dynamical geometry, and show how the state can be reconstructed from machine records. The few-qubit analysis shows how these tools can be used. For three qubits, relations between probabilities explain regular families and allow us to compare the emergence of islands and unstable regions as the organization of Born weights changes. The lift then distinguishes states with identical probabilities and reveals phase relations conditioned on the third qubit. The joint parameterization of Born probabilities and phases is thereby expressed in geometric structures that can be followed across a family.

Section~\ref{sec:theory} develops the CCBM. Section~\ref{sec:results} studies Born geometry: the two-qubit reduction, conditions for regularity in three qubits, and numerical comparisons between families. Section~\ref{sec:phasegeometry} examines their phases, and Sec.~\ref{sec:reconstruction} tests reconstruction from records. We then discuss the scope of the representation. The appendices collect its connection to RBMs, inversion formulas, expectation values, and perturbation dynamics.

\section{From Suzuki's machine to quantum-state representation}\label{sec:theory}
\subsection{Quantum state and complex parameterization}
Consider a pure state of $n$ qubits,
\begin{equation}
 |\psi\rangle=\sum_{\bs\in\{0,1\}^n}C_{\bs}|\bs\rangle,
 \qquad \sum_{\bs}|C_{\bs}|^2=1,
 \label{eq:state}
\end{equation}
where $|\bs\rangle$ are computational-basis states and $\bs=(s_1,\ldots,s_n)$ is a binary configuration. The complex coefficients $C_{\bs}$ can be written in terms of magnitude and phase as
\begin{equation}
 C_{\bs}=\sqrt{p_{\bs}}\,e^{\iu\phi_{\bs}},\qquad
 p_{\bs}=|C_{\bs}|^2.
 \label{eq:polar}
\end{equation}
The probabilities $p_{\bs}$ correspond to measurements in the computational basis, and $\phi_{\bs}$ are relative phases, with $\phi_{\bm0}=0$ as the reference. Throughout this work we assume $p_{\bs}>0$ for every configuration.

The CCBM is parameterized by a complex function of binary variables. Its real and imaginary parts specify the wave-function magnitudes and phases, respectively. Once these parameters are chosen, the rules developed below assign motion in the hypercube and a complex lift to the state. The argument of this lift provides the angular coordinate wherever it is defined.

To express dependence between visible variables, define $L(\bs)$ through $e^{L(\bs)}=C_{\bs}/C_{\bm0}$, with $L(\bm0)=0$. After choosing a logarithm branch for each configuration, its general binary expansion is
\begin{align}
 L(\bs)={}&\sum_i a_i s_i+\sum_{i<j}J_{ij}s_i s_j\nonumber\\
 &+\sum_{i<j<k}K_{ijk}s_i s_j s_k+\cdots .
 \label{eq:loggeneral}
\end{align}
The coefficients $a_i$, $J_{ij}$, and $K_{ijk}$ are complex and describe one-, two-, and three-body contributions to the log amplitude. The expansion ends at order $n$. For two qubits, $a_i$ and $J_{12}$ suffice; a general three-qubit state also requires $K_{123}\equiv K$. Their real parts determine magnitudes and their imaginary parts determine phases:
\begin{equation}
 p_{\bs}=\frac{e^{2\operatorname{Re}L(\bs)}}{
 \sum_{\br}e^{2\operatorname{Re}L(\br)}},\qquad
 \phi_{\bs}=\operatorname{Im}L(\bs)\pmod{2\pi}.
 \label{eq:logpolar}
\end{equation}

Equation~\ref{eq:loggeneral} extends the usual pairwise Boltzmann machine in two ways: its parameters are complex, and it includes explicit three-body couplings $K_{ijk}$ and interactions of higher order. Retaining all orders gives a complete representation of any pure state with nonzero computational-basis coefficients. Arbitrary correlations between qubits can therefore be encoded directly in these visible couplings, without introducing hidden layers or requiring a deep Boltzmann machine. With normalization and global phase fixed, the state has $2(2^n-1)$ independent real parameters. The families studied here impose relations among them to explore specific Born and phase structures. Different logarithm branches may change the written parameters, but the probabilities and factors $e^{\iu\phi_{\bs}}$ determine the same representation.

The coefficients can also be obtained from a complex RBM by summing over its hidden variables. Appendix~\ref{app:rbm} explains this connection and distinguishes its weights from the visible couplings in Eq.~\eqref{eq:loggeneral}. An RBM is one possible input parameterization; the CCBM rules are defined independently of it.

\subsection{Deterministic dynamics associated with Born probabilities}
The classical Boltzmann model with pairwise interactions assigns each configuration the energy and probability
\begin{align}
 E(\bs)&=-\sum_i b_i s_i-\sum_{i<j}w_{ij}s_i s_j,\nonumber\\
 p(\bs)&=Z^{-1}e^{-E(\bs)/T},\qquad
 Z=\sum_{\bs}e^{-E(\bs)/T}.
 \label{eq:classical}
\end{align}
The biases $b_i$ and couplings $w_{ij}$ set individual preferences and interactions between variables; $T>0$ sets the energy scale, and $Z$ normalizes the distribution. The pairwise restriction belongs to this classical reference model. The Born probabilities used here may contain the higher-order contributions in Eq.~\eqref{eq:loggeneral}.

Suzuki associated deterministic dynamics in the unit hypercube with the classical distribution~\cite{Suzuki2013CBM,Suzuki2013Billiards}. The dynamical state consists of a position $\bx=(x_1,\ldots,x_n)\in[0,1]^n$ and a label $\bs$. Each bit determines the direction of motion of its coordinate: positive for $s_i=0$ and negative for $s_i=1$. Each label is associated with a complete copy of the hypercube.

Let $\bs^{(i)}$ denote the configuration obtained by flipping bit $i$, and let $\bs_{-i}$ denote the remaining bits. The conditional probability of the current value of $s_i$ is
\begin{equation}
 p(s_i\mid\bs_{-i})=
 \frac{p_{\bs}}{p_{\bs}+p_{\bs^{(i)}}}.
 \label{eq:conditional}
\end{equation}
The speed of each coordinate is inversely proportional to this probability:
\begin{align}
 \dot x_i=v_i(\bs)
 &=\frac{c_i(1-2s_i)}{p(s_i\mid\bs_{-i})}\nonumber\\
 &=c_i(1-2s_i)\left(1+\frac{p_{\bs^{(i)}}}{p_{\bs}}\right),
 \label{eq:flow}
\end{align}
where $c_i>0$ sets the speed scale. The factor $1-2s_i$ determines the sign of the velocity, while the probability ratio determines its magnitude. A configuration that is less probable than its neighbor has a higher speed in the coordinate connecting them.

During each flight the label is fixed and motion is uniform along a straight line. When $x_i$ reaches 1, we set $s_i=1$; when it reaches 0, we set $s_i=0$. Position is continuous, and all velocity components are recalculated using the new label. Reflection is therefore generally nonspecular. We consider trajectories with isolated impacts and exclude simultaneous arrivals at multiple faces.

The CCBM base dynamics is obtained by inserting the Born probabilities into Eq.~\eqref{eq:flow}. Because this rule depends only on conditional probabilities, it also applies to distributions with higher-order interactions. We set $c_i=1$ for all coordinates in the numerical comparisons.

\subsection{Invariant density and Born distribution}
Consider an ensemble of particles in which the fraction with label $\bs$ is $p_{\bs}$ and positions are uniformly distributed within each label. The corresponding joint measure is
\begin{equation}
 \mu(\bs,\dd^n x)=p_{\bs}\,\dd^n x,
 \qquad \dd^n x=\dd x_1\cdots\dd x_n.
 \label{eq:mu}
\end{equation}
Here $\dd^n x$ is a hypervolume element, and $\mu(\bs,\dd^n x)$ is the probability of finding a particle in it with label $\bs$. The density $p_{\bs}$ is constant in position but depends on the configuration. Since the hypercube has unit volume, integrating over positions recovers the Born probability $p_{\bs}$.

Invariance of this measure requires preserving the density during both flights and impacts. In the interior, constant velocity translates the distribution without compressing it. At a face, the normal particle flux per unit area is the density times $|v_i|$. During a time $\dd t$, the fraction reaching a face element $\dd A$ is $p_{\bs}|v_i|\dd A\dd t$. The velocities in Eq.~\eqref{eq:flow} satisfy
\begin{equation}
 p_{\bs}|v_i(\bs)|=
 c_i(p_{\bs}+p_{\bs^{(i)}})=
 p_{\bs^{(i)}}|v_i(\bs^{(i)})|.
 \label{eq:flux}
\end{equation}
The fluxes before and after a label change are equal. This balance between density and speed prevents probability from accumulating at the face and preserves the uniform distribution within each configuration. The machine's stationary ensemble therefore has the Born distribution as its discrete marginal.

Measure preservation is an ensemble property. Agreement between its weights and the time fractions of an individual trajectory additionally requires that the trajectory sample the corresponding distribution.

\subsection{Unfolded representation and unit cell}\label{sec:unfolded}
A useful dynamical representation must distinguish states that share a position but have different velocities, as phase space does in classical mechanics. Here, position alone does not determine the motion: the binary configuration is also needed. For three qubits, a given wall has four possible outgoing configurations, so plotting returns without distinguishing their labels would superimpose four different sections. Unfolding assigns each configuration to a separate cell and each outgoing label to a separate region of the section, allowing the dynamics to be visualized without this ambiguity.

Impacts can be visualized as crossings between cells through the coordinate transformation
\begin{equation}
 Q_i=2m_i+2s_i+(1-2s_i)x_i,\qquad m_i\in\mathbb{Z}.
 \label{eq:unfolded}
\end{equation}
For $s_i=0$, the coordinate $Q_i$ traverses $[2m_i,2m_i+1]$ as $x_i$ increases from zero to one. For $s_i=1$, it traverses $[2m_i+1,2m_i+2]$ as $x_i$ returns from one to zero. Completing this path increments $m_i$. Thus, $Q_i$ is continuous at each impact and obeys $\dot Q_i=|v_i(\bs)|>0$ during flights.

\begin{figure*}[t]
\centering\includegraphics[width=\textwidth]{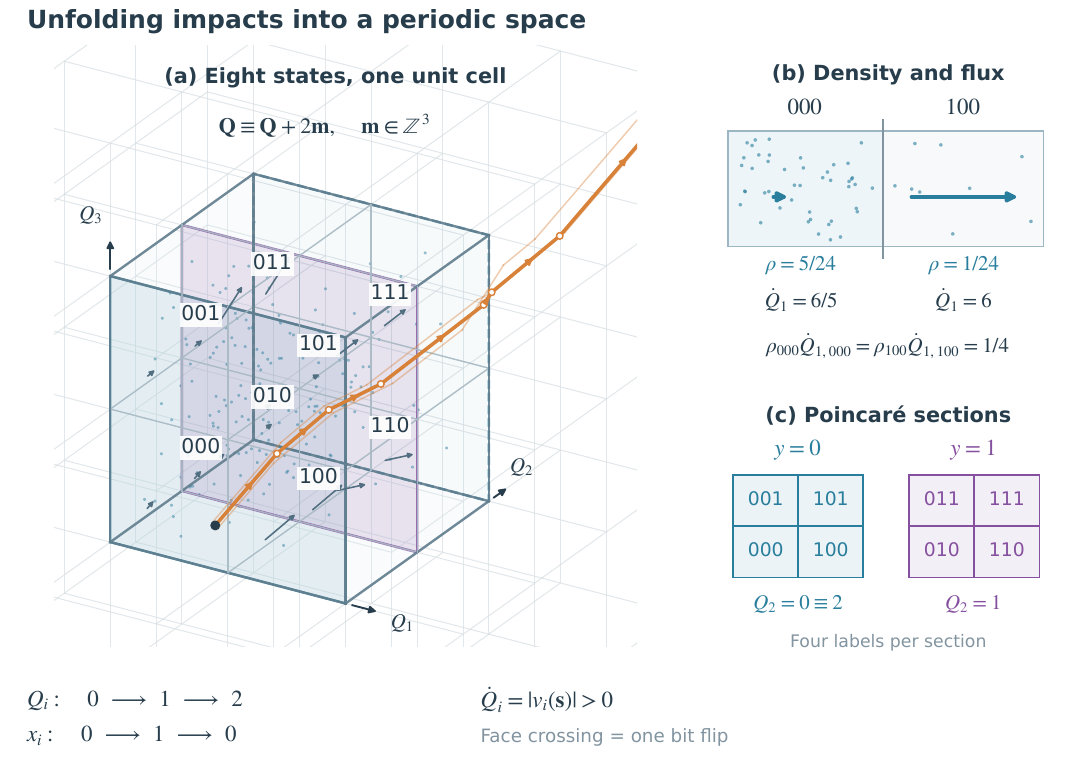}
\caption{Periodic unfolding of the three-qubit dynamics. (a) The unit cell contains eight cubes, one per configuration. Arrows show the constant field inside each cube; orange trajectories are calculated using Suzuki's rule and continue into neighboring cells. The blue cloud illustrates the stationary ensemble, with uniform positions and point counts proportional to Born probabilities; it is not obtained by propagating an ensemble launched from a face. We use $p_{000}=p_{001}=p_{010}=p_{101}=5/24$ and $1/24$ for the remaining configurations, with $c_i=1$. (b) Density and normal speed balance exactly at the interface $000\to100$. (c) The faces $y=0$ and $y=1$ are represented by the blue and purple planes, respectively, each with four outgoing labels. The squares indicate section regions, not return plots.}
\label{fig:unfolded}
\end{figure*}

In three dimensions, the eight configurations occupy the unit cubes of a cell of side length two, repeated periodically in all three directions (Fig.~\ref{fig:unfolded}). Inside each cube, $s_i=\lfloor Q_i\rfloor\bmod2$ and the velocity field is constant. Each face crossing flips one bit and changes the velocity; the trajectory continues in the neighboring cube. Reaching $111$ from $000$ requires at least three crossings for the isolated impacts considered here.

The absolute value of the coordinate transformation's Jacobian determinant is one. The stationary measure therefore retains the density $\rho(\mathbf Q)=p_{\bs}$ within each cube, normalized over one unit cell. Densities may differ between cubes. The flux equality in Eq.~\eqref{eq:flux} balances these differences at their interfaces. Together with uniform motion inside each cell, this condition explains Born invariance in the unfolded representation.

Sections on the physical faces $y=x_2=1$ appear on the planes $Q_2=1\pmod2$, while those on $y=0$ appear on $Q_2=0\pmod2$. In the cell $[0,2]^3$, the first plane is internal and the second lies on the identified periodic faces. Each plane is divided into four squares according to the other two bits. Labels are taken immediately after crossing, as in the sections of Sec.~\ref{sec:sections}.

\subsection{Phase field and angular lift}
We incorporate phases by assigning the unit factor $u_{\bs}=e^{\iu\phi_{\bs}}$ to each vertex $\bs$ and interpolating linearly in each coordinate:
\begin{equation}
 B_{\bs}(\bx)=\prod_{i=1}^{n}x_i^{s_i}(1-x_i)^{1-s_i},
 \qquad U(\bx)=\sum_{\bs}B_{\bs}(\bx)u_{\bs}.
 \label{eq:U}
\end{equation}
The weights $B_{\bs}$ are nonnegative and sum to one, so $U$ is a convex combination of phase factors. Each vertex recovers its assigned value, $U(\bs)=u_{\bs}$. This interpolation preserves phase equivalence modulo $2\pi$ and depends only on phases, since it uses unit factors rather than full amplitudes.

The choice of this field sets the rules for the angular component. We require it to reproduce the vertex factors and to be affine in each coordinate with the others held fixed. Within this class, the vertex values uniquely determine Eq.~\eqref{eq:U}: interpolation between the two faces of one coordinate is applied successively to the others. The affine condition is a choice of representation. The field introduces no additional free parameters beyond the state phases.

Restriction to a face uses only its vertices. For a bit $s_i=b$, with $b=0,1$, we write explicitly
\begin{equation}
 U(\bx)\big|_{x_i=b}=\sum_{\bs:s_i=b}
 \left[\prod_{j\ne i}x_j^{s_j}(1-x_j)^{1-s_j}\right]u_{\bs}.
 \label{eq:facefield}
\end{equation}
This property associates faces with well-defined subsets of coefficients and will allow their phases to be compared. Coordinate permutations and inversions $x_i\mapsto1-x_i$, accompanied by the same operations on binary indices, transform the field in the corresponding way. The angular rule thus respects the cube symmetries used to organize Born probabilities.

A global phase multiplies $U$ by the same factor and shifts all its angles by a constant. If the phase factors separate between two groups of qubits, $u_{\bs_A\bs_B}=u^A_{\bs_A}u^B_{\bs_B}$, the sum also separates and $U=U_AU_B$. These properties allow phase references and independent contributions to be interpreted. A local phase transformation of a general state can change interior heights; the invariant phase combinations studied below are defined directly from its coefficients.

We first define the complex lift of a base trajectory by
\begin{equation}
 \zeta(t)=U(\bx(t)),\qquad
 \dot\zeta=\nabla U(\bx)\cdot\bv(\bs).
 \label{eq:complexlift}
\end{equation}
The complex variable $\zeta$ records the field along the motion, with initial value $\zeta(0)=U(\bx(0))$. This condition preserves the equality during each flight. At impacts, $\zeta$ remains continuous and its derivative is recalculated with the new velocity. Since $U$ is a polynomial in the coordinates, both $\zeta$ and its derivative during a flight remain defined at $U=0$. The lifted motion $(\bx,\bs,\zeta)$ therefore continues through these points without an additional jump rule.

Where $U\ne0$, the angle is a coordinate of this lift and evolves according to
\begin{equation}
 e^{\iu\theta(\bx)}=\frac{U(\bx)}{|U(\bx)|},\qquad
 \dot\theta=\operatorname{Im}\frac{\nabla U\cdot\bv(\bs)}{U}.
 \label{eq:theta}
\end{equation}
The product $\nabla U\cdot\bv$ is the field derivative along the trajectory. The imaginary part of its ratio to $U$ is the angular velocity. At impacts, $\theta$ remains continuous if $U\ne0$, although its derivative may change. Sections are constructed by evaluating the angle at return points.

The angular representation uses $(\bx,\theta)$, with $\theta$ defined modulo $2\pi$ and label $\bs$ associated with each flight. Both $\zeta$ and $\theta$ are determined by position: they introduce no independent degrees of freedom and do not modify the velocities. Varying phases at fixed probabilities therefore leaves the hypercube motion unchanged while modifying its angular representation. The values $\phi_{\bs}$ belong to the vertices; the field $\theta(\bx)$ interpolates their information at interior positions wherever it is defined.

The interpolation may have zeros due to cancellation of complex factors, even when every state coefficient is nonzero. On this nodal set, $\zeta=0$ remains defined, but its argument does not exist. Equation~\eqref{eq:theta} applies on intervals avoiding zeros; after a crossing, the field argument is evaluated again without imposing an angular continuation through the node. The full crossing is represented using $\operatorname{Re}\zeta$ and $\operatorname{Im}\zeta$. A jump or rapid angular variation near a zero does not imply instability of the base motion.

If all phase factors lie in an open semicircle, $U$ cannot vanish: after a common rotation, its real part is strictly positive. Other choices produce nodal sets, whose geometry is examined in the few-qubit examples and in Sec.~\ref{sec:nodalcrossing}. On an invariant region bounded away from these zeros, the angular lift is smooth and preserves the regularity of the base motion.

Equations~\eqref{eq:flow} and \eqref{eq:complexlift}, with $U$ given by Eq.~\eqref{eq:U}, the stated initial condition, and the impact rule, define the CCBM for the trajectories with isolated impacts considered here. Equation~\eqref{eq:theta} gives its angular representation outside the nodal set. The word complex refers to the representation of magnitudes and phases through complex parameters; spatial velocities remain real. The name retains that of Suzuki's machine, although some families have regular dynamics. Other interpolations could reproduce the same vertex phases while producing different interior geometries. Here we fix the rule that is affine in each coordinate, with the properties above, to compare states within a single representation.

\subsection{Correspondence with the state}
Once the basis, speed scales $c_i$, and interpolation are fixed, the complex parameters determine a CCBM and, through Eq.~\eqref{eq:logpolar}, a normalized wave function. Conversely, the velocities give the ratios $p_{\bs^{(i)}}/p_{\bs}$, and the angular field gives the vertex phases. Ratios along a set of edges connecting all configurations, together with normalization, determine the probabilities. The vertex factors determine the relative phases. The full representation therefore preserves the state coefficients up to a global phase.

This correspondence concerns the complete machine rules. A finite set of returns may contain only part of that information. Section~\ref{sec:reconstruction} examines the conditions under which it can be recovered from records. The stationary measure, section geometry, and angular field are different aspects of the same representation, each interpreted according to its definition.

\subsection{Poincar\'e sections}\label{sec:sections}
Poincar\'e sections provide a reduced description of the structures that organize the motion. On a transverse section, invariant structures of the flow are represented by invariant sets of the return map. In our three-dimensional base dynamics, periodic orbits give finite cycles, invariant tori appear as curves, and chaotic regions are explored by irregular sequences of returns. The geometry and stability of these sets reveal how regular and chaotic motion are organized. For trajectories that repeatedly cross the section, the return map, together with the return times and the known flow between crossings, describes their continuous-time dynamics.

We choose a face $x_i=0$ and record returns immediately after impact. The outgoing label satisfies $s_i=0$, while the remaining bits provide $2^{n-1}$ possibilities. For each label we define the section
\begin{equation}
 \Sigma_{\bs}^{(i)}=
 \{(\bx,\bs):x_i=0,\ 0<x_j<1\ (j\ne i)\}.
 \label{eq:section}
\end{equation}
Returns are separated by label to retain a single velocity branch in each section. Where $U\ne0$, the angular coordinates are $(\bx_{-i},\theta)$, where $\bx_{-i}$ collects the coordinates tangent to the face. At a nodal return we retain the complex record $\zeta=0$ and the position, without assigning an angular height.

The stationary crossing distribution is obtained by weighting the density by the normal speed. For the outgoing label $\bs$, the flux is $p_{\bs^{(i)}}|v_i(\bs^{(i)})|=c_i(p_{\bs}+p_{\bs^{(i)}})$. Summing over all labels with $s_i=0$ gives $c_i$, so the normalized crossing measure is
\begin{equation}
 \dd\nu_{\bs}^{(i)}=
 (p_{\bs}+p_{\bs^{(i)}})\,\dd^{n-1}x_{-i},
 \qquad s_i=0.
 \label{eq:sectionmeasure}
\end{equation}
The element $\dd^{n-1}x_{-i}$ is the tangential area of the section, and $\dd\nu_{\bs}^{(i)}$ is the fraction of crossings through it with that label. Within each label the measure is uniform in area; its weight is the sum of the two probabilities connected by the impact. The impact distribution and the Born distribution therefore carry different weights.

A section constructed from a finite collection of trajectories shows the regions visited by that collection. Regular curves and gaps reflect the organization of motion and do not directly identify concentrations or zeros of Born probabilities.

\subsection{Estimation from occupation times}\label{sec:occupation}
Probabilities can be estimated by accumulating the time spent by trajectories in each configuration. If $\tau_{ak}$ is the recorded duration of flight $k$ of trajectory $a$, and $\bs_{ak}$ is its label, the estimator is
\begin{equation}
 \widehat p_{\bs}=
 \frac{\displaystyle\sum_{a,k}\tau_{ak}
 \mathbf 1_{\{\bs_{ak}=\bs\}}}
 {\displaystyle\sum_{a,k}\tau_{ak}}.
 \label{eq:occupation}
\end{equation}
The indicator function $\mathbf 1_{\{\bs_{ak}=\bs\}}$ selects flights in the configuration of interest. The numerator sums their durations, and the denominator is the total observation time. Flights crossing the endpoints of the observation interval contribute only their recorded durations. This weighting distinguishes time occupation from impact frequency.

Convergence to Born probabilities requires sampling the measure in Eq.~\eqref{eq:mu}: it may be obtained as a time average along an ergodic trajectory or as an average over independent realizations of the stationary ensemble, observed for the same interval. A collection launched from a face does not necessarily satisfy this condition, even if its size or propagation time increases. Regular invariant regions may retain different occupation fractions.

Probabilities can also be recovered directly from velocities, which encode ratios between neighboring configurations. Appendix~\ref{app:inverse} develops this algebraic alternative, used in reconstruction tests with short records. This separates the information contained in the dynamical rule from the statistical convergence of a time average.

\section{Born geometry for two and three qubits}\label{sec:results}
\subsection{Two qubits: analytical structure of the motion}\label{sec:two}
We start with a state $|\psi\rangle=C_{00}|00\rangle+C_{01}|01\rangle+C_{10}|10\rangle+C_{11}|11\rangle$ with four nonzero coefficients. The particle now moves in a square, with velocities determined by the four probabilities $p_{ij}=|C_{ij}|^2$. We use equal scales in both coordinates, $c_1=c_2=1$, to compare states.

To establish the two-qubit reference, we use Suzuki's reduction~\cite{Suzuki2014Two}. Let $p_i(s_i)$ be the probability of bit $i$ after summing over the other bit. For example, $p_1(0)=p_{00}+p_{01}$. Define
\begin{equation}
 q_i=(1-2s_i)p_i(s_i)x_i.
 \label{eq:q}
\end{equation}
When $s_i=0$, $q_i$ traverses $[0,p_i(0)]$; when $s_i=1$, it traverses $[-p_i(1),0]$. The total length is one. We identify the endpoints $-p_i(1)$ and $p_i(0)$ so that each coordinate describes a circle. An impact at $x_i=1$ crosses this identification, while an impact at $x_i=0$ crosses $q_i=0$. Together, the two circles form a torus.

Differentiating Eq.~\eqref{eq:q} and substituting the velocities gives
\begin{equation}
 \dot q_1=\dot q_2=
 \frac{p_1(s_1)p_2(s_2)}{p(s_1,s_2)}>0.
 \label{eq:qdot}
\end{equation}
Both coordinates always advance at the same speed. This speed can change between configurations, but the difference between $q_1$ and $q_2$ is preserved. Introducing a time $\tau$ that absorbs the common factor gives
\begin{equation}
 \frac{\dd q_1}{\dd\tau}=\frac{\dd q_2}{\dd\tau}=1,
 \qquad q_1-q_2\pmod 1=\text{constant}.
 \label{eq:torus}
\end{equation}
After advancing one unit in $\tau$, both coordinates complete a full turn. Trajectories with isolated impacts are therefore periodic for any positive two-qubit distribution when the scales are equal. Correlations change the orbit's shape in the square and its flight times, but do not generate chaos in this case.

The period in the original time $t$ is obtained by summing the time spent in the configurations visited. With $d=q_1-q_2$ and $q_1=a+\tau$, it can be written as
\begin{equation}
 T(d)=\int_0^1
 \frac{p(s_1(a+\tau),s_2(a+\tau-d))}
 {p_1(s_1(a+\tau))p_2(s_2(a+\tau-d))}\,\dd\tau.
 \label{eq:period}
\end{equation}
Here $s_i(q_i)$ identifies which of the two intervals of coordinate $q_i$ contains the point. The integrand is constant on each segment with a fixed configuration. Calculating the period therefore amounts to summing the lengths of these segments, weighted by the factor converting $\tau$ into $t$.

For unequal scales, motion on the torus obeys $\dd q_i/\dd\tau=c_i$. If their ratio is irrational, the orbit does not close, as in the sampling condition studied by Suzuki~\cite{Suzuki2014Two}. We retain equal scales throughout this work so that observed differences can be attributed to the represented wave function.

Periodicity illustrates the distinction between invariance and time sampling. For the uniform distribution $p_{s_1s_2}=1/4$, the trajectory starting at $(0.2,0.6)$ with label $00$ spends fractions $(0.3,0.2,0.2,0.3)$ of each period in configurations $(00,01,10,11)$. These fractions persist as the orbit repeats, although the machine's uniform measure has marginal $p_{s_1s_2}=1/4$.

\subsection{Three-qubit families with two distinct Born probabilities}
The CCBM assigns dynamics to each choice of state parameters. To explore this representation, we first compare families that preserve coefficient magnitudes while changing their organization in the basis. We then study profiles with eight distinct probabilities. Phases remain fixed during this analysis of the base geometry.
To separate probability concentration from assignment to the computational basis, we choose a set $A$ of $m$ favored configurations and define
\begin{align}
 p_A(\bs;r)&=\frac{1+(r-1)\mathbf1_A(\bs)}{mr+8-m},\nonumber\\
 |\psi_A\rangle&=\sum_{\bs}\sqrt{p_A(\bs;r)}\,|\bs\rangle,
 \qquad m=2,3,4,\quad r\ge1.
 \label{eq:bornfamily}
\end{align}
The indicator $\mathbf1_A$ is one on the selected configurations and zero elsewhere. The parameter $r$ is the ratio between the probability of each favored configuration and that of each remaining configuration: $p_{\mathrm{fav}}/p_{\mathrm{rest}}=r$. Thus, $r=1$ gives the uniform distribution, while increasing $r$ increases the contrast between the two groups. At fixed $m$ and $r$, the eight probability values and their Shannon entropy are preserved. Only the coefficient receiving each weight changes. Phases are initially set to zero.

The eight configurations label the cube's vertices. Swapping qubits permutes its axes, while interchanging 0 and 1 for a qubit reflects one axis. These operations generate the cube's 48 symmetries. Arrangements related by these transformations belong to the same class: at fixed $r$ and equal speed scales, their dynamics are equivalent after transforming the coordinates and observation face. The 28 arrangements of two favored vertices form three classes, distinguished by the number of differing bits (the Hamming distance). The 56 arrangements of three favored vertices give three classes, and the 70 arrangements of four give six, with representatives listed in Table~\ref{tab:classes}.

\begin{figure*}[tp]
\centering\includegraphics[width=.92\textwidth]{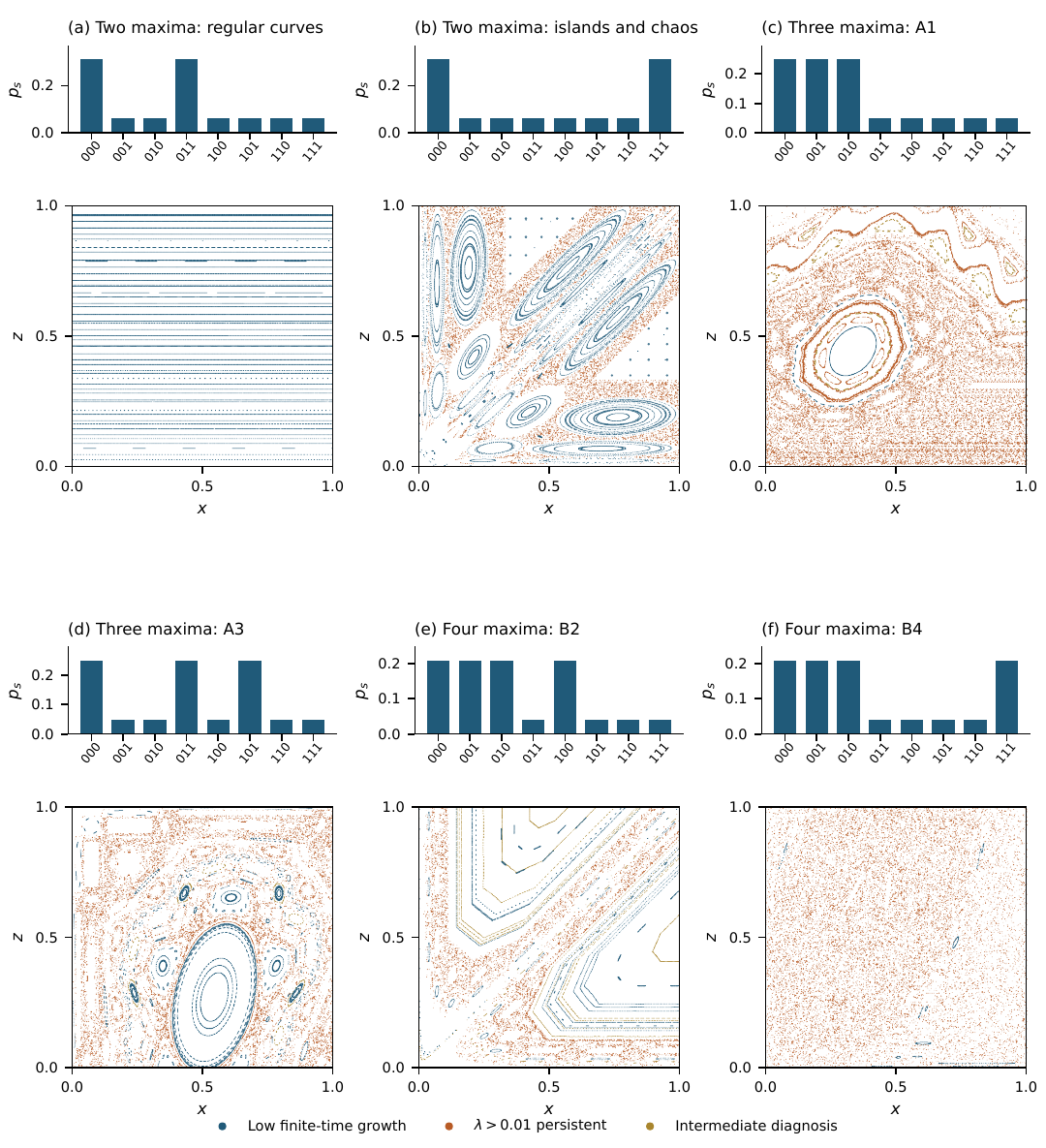}
\caption{Born organization and the geometry of regular and unstable regions at $r=5$. Each panel shows probabilities above and returns to $y=0$ with label $000$ below. Each pair preserves the same probability values. (a),(b) Two maxima at Hamming distances two and three: regular curves versus islands surrounded by unstable trajectories. (c),(d) Three maxima in classes A1 and A3: the position and visible extent of islands change, as does the distribution of unstable returns around them. (e),(f) Four maxima in B2 and B4: unstable regions between regular curves versus a section extensively visited by unstable trajectories, with small visible islands. Panels (a),(c),(e),(f) retain the original 50 trajectories. Panel (b) adds 12 periodic trajectories and four trajectories within regular islands, together with the $x\leftrightarrow z$ images of all 66 trajectories; panel (d) adds six trajectories resolving small island chains. Returns for $20<t\le800$ are shown. The isolated blue dots in the triangular regions of (b) are repeating periodic returns. The added and symmetry-related trajectories are used for visualization only, not for flux estimates. Orange: rates above 0.01 at 3200 and 12800; blue: final rate with magnitude below 0.002; ochre: intermediate diagnosis. Regularity in (a) is also established analytically. Colors are determined by tangent growth, not point density; unclassified empty regions are not automatically identified as regular.}
\label{fig:bornorganization}
\end{figure*}

Figure~\ref{fig:bornorganization} compares the resulting sections. For $A=\{000,001\}$, the third bit is independent of the other two and equally likely to be 0 or 1. Hence $p(\bs)=p_{xy}(s_x,s_y)/2$, where $p_{xy}$ is the joint probability of the first two bits. The motion combines regular two-qubit dynamics with periodic motion in $z$. For $A=\{000,011\}$ [panel (a)], regularity instead follows from equal speed magnitudes in $y$ and $z$, as explained below. Opposite favored vertices, $A=\{000,111\}$, give regular islands alongside unstable trajectories [panel (b)].

For this last family, define $\sigma_i=1-2s_i$: it is $+1$ for bit 0 and $-1$ for bit 1. Writing $g=\log r/4$ gives the symmetric distribution
\begin{equation}
 p(\bs)\propto e^{g(\sigma_x\sigma_y+\sigma_x\sigma_z+\sigma_y\sigma_z)},
 \qquad g=\frac{\log r}{4},\quad\sigma_i=1-2s_i.
 \label{eq:symmetricborn}
\end{equation}

The parameter $g$ expresses the same probability contrast in exponential form: the favored configurations have weights $e^{4g}=r$ times those of the others. Averages with Born probabilities give $\langle\sigma_i\rangle=0$ and $\langle\sigma_i\sigma_j\rangle=(r-1)/(r+3)$ for every pair $i\ne j$. Increasing these correlations does not necessarily increase instability.

\subsection{Probability relations that preserve regular motion}
A sufficient condition for regularity is separation of one qubit from the other two, $p(\bs)=p_i(s_i)p_{jk}(s_j,s_k)$. The period of the two-qubit subsystem may depend on the orbit, while the coordinate of the separated qubit completes one cycle per unit time. Their combined motion is periodic or quasiperiodic.

Regularity can also occur when the probabilities depend on all three bits. Consider a distribution that depends on $s_x$ and on whether $s_y$ and $s_z$ are equal, but does not distinguish $00$ from $11$ or $01$ from $10$ for this pair. We describe it using two binary labels: $a=s_x$ and $e=s_y\mathbin{\oplus}s_z$, where $\oplus$ denotes addition modulo two, so $e=0$ for equal bits and $e=1$ for unequal bits. Let $P_{ae}>0$ be the joint probability of these labels, with $\sum_{a,e}P_{ae}=1$. Each choice of $(a,e)$ corresponds to two equally probable configurations of the three qubits; their individual Born probabilities are therefore
\begin{equation}
 p(\bs)=\frac{P_{ae}}2.
 \label{eq:lockedfamily}
\end{equation}

Flipping either $s_y$ or $s_z$ changes $e$ in the same way. The two flips therefore lead to the same probability ratio in the velocity rule, giving $|v_y|=|v_z|$ at every configuration. To follow this constraint through the reflections, we use the unfolded coordinates introduced in Sec.~\ref{sec:unfolded}: $u_i=x_i$ for $s_i=0$ and $u_i=2-x_i$ for $s_i=1$, with values identified modulo two. They advance with $\dot u_i=|v_i|$, so the separation $\delta=u_z-u_y\pmod2$, chosen in $[0,2)$, remains constant. Thus, the motion in $y$ and $z$ is linked rather than independent.

This conserved separation allows us to reduce the return dynamics to a single coordinate. Define $A_a=\sum_eP_{ae}$ and $B_e=\sum_aP_{ae}$, the probabilities of the labels $a$ and $e$, respectively. The velocity rule gives $\dd u_x/\dd u_y=B_e/A_a$. We absorb the denominator by rescaling the two halves of the unfolded $x$ coordinate according to $A_0$ and $A_1$:
\begin{equation}
 \xi_x(u_x)=\begin{cases}A_0u_x,&0\le u_x\le1,\\
 A_0+A_1(u_x-1),&1\le u_x\le2,
 \end{cases}
\end{equation}

Since $A_0+A_1=1$, $\xi_x$ runs around a circle of unit length. We record its value whenever $u_y$ completes a full turn, that is, advances by two. The next recorded value, denoted by $\xi_x'$, is
\begin{align}
 \xi_x'&=\xi_x+\rho(\delta)\pmod1,\nonumber\\
 \rho(\delta)&=2B_0|1-\delta|+2B_1[1-|1-\delta|].
 \label{eq:rotationgeneral}
\end{align}

Here $\rho(\delta)$ is the advance of $\xi_x$ between consecutive returns. During one turn of $u_y$, the bits $s_y$ and $s_z$ are equal over a total coordinate interval $2|1-\delta|$ and unequal over the remaining interval. Integrating $\dd \xi_x/\dd u_y=B_e$ over these two intervals gives Eq.~\eqref{eq:rotationgeneral}. Because $\delta$ is conserved, every return adds the same amount: the map is a rigid rotation. Rational values of $\rho$ give repeating return points, whereas irrational values give quasiperiodic motion. Between nearby surfaces with different $\delta$, separations can grow at most linearly with the number of returns. Together with bounded return times for positive probabilities, this excludes exponential growth for trajectories with isolated impacts. Numerical checks on twenty random probability tables reproduce this map with errors below $1.1\times10^{-14}$.

The same mechanism explains several regular families with four favored configurations. Select a subset $I$ of the three bits and assign the larger probability to configurations with an even number of ones among those bits. This condition selects four of the eight configurations. Using $\sigma_i=1-2s_i$, the product $\chi_I=\prod_{i\in I}\sigma_i$ equals $+1$ for even parity and $-1$ for odd parity. The distribution is
\begin{equation}
 p(\bs)=\frac{1+\eta\chi_I(\bs)}8,
 \qquad\eta=\frac{r-1}{r+1},\quad
 \chi_I=\prod_{i\in I}\sigma_i.
 \label{eq:parityborn}
\end{equation}

The parameter $\eta$ sets the contrast: the two probabilities are $(1+\eta)/8$ and $(1-\eta)/8$, whose ratio is $r$. Selecting one, two, or three bits gives, respectively, a face of the cube, two opposite edges, or the four vertices of a parity tetrahedron. For any coordinate $i\in I$, flipping its bit changes the parity, giving $\dot u_i=2/(1+\eta\chi_I)$. Coordinates outside $I$ have speed two because flipping their bits leaves the probability unchanged. All coordinates in $I$ therefore advance at the same speed and retain their mutual separations. With one or three selected bits, shifting these coordinates by one reverses the parity; the times for the two halves of a full turn consequently sum to one, and the motion repeats. With two selected bits, the regular two-qubit motion combines with the periodic motion of the remaining coordinate. These three classes account for 14 of the 70 arrangements of four favored configurations.

Regularity therefore depends on which configurations share a probability, not simply on how many are favored. For example, start with $A=\{000,001\}$, then add $110$, and finally $111$. At $r=5$, this sequence goes from regular motion to the predominantly unstable class A2 and back to the regular class B5 (Table~\ref{tab:classes}). In the final set, the favored configurations are precisely those with $s_x=s_y$, independently of $s_z$: adding the fourth configuration completes the equality condition that preserves regularity. In a fully periodic case, each trajectory contributes only a repeating finite set of section points, but superimposing many initial conditions can still produce a cloud. The figures therefore distinguish trajectories by tangent growth, rather than identifying chaos from point density alone.

\subsection{Extent of unstable regions}

To distinguish unstable trajectories from regular ones, we measure how a small perturbation to the initial position grows along each trajectory. We use the finite-time Lyapunov growth rate $\lambda_T=T^{-1}\log[\|\delta\bm x(T)\|/\|\delta\bm x(0)\|]$, where $\delta\bm x$ is an infinitesimal displacement evolved by the linearized dynamics and $T$ is the observation time. Here time zero denotes the start of measurement after an initial interval of duration 20; $T$ is the subsequent observation time. A positive value indicates net amplification over that interval; persistent positive values at increasing $T$ provide numerical evidence of exponential separation. The calculation follows each flight directly to the next wall, without time discretization, and updates the perturbation at the impact, as described in Appendix~\ref{app:stability}. We classify a trajectory as showing persistent growth when $\lambda_T>0.01$ at both of the two longest observation times. If $|\lambda_{T_{\max}}|<0.002$ at the final time $T_{\max}$, we classify it as low growth; the remaining cases are intermediate. These thresholds use the time units of our dynamics. This is a finite-time diagnosis, not a determination of the infinite-time Lyapunov exponent; the two observation intervals start from the same initial condition and overlap.

We next ask how frequent unstable trajectories are for each Born distribution. A single section may emphasize a particular island or unstable region, so we sample all six faces of the cube, with positions distributed uniformly within each of their four outgoing configurations. Let $f_{i,w,s}^{(\tau)}$ be the fraction of these initial conditions whose trajectories exceed the growth threshold $\tau$ at both observation times, for face $x_i=w$ and outgoing configuration $s$. Here $i$ specifies the coordinate normal to the face and $w=0$ or $1$ specifies the wall. The four configurations do not generally carry equal stationary crossing rates: Sec.~\ref{sec:sections} gives their weights as $p_s+p_{s^{(i)}}$, where $s^{(i)}$ differs from $s$ only in bit $i$. We therefore combine the measured fractions using these crossing weights, then average over the six faces:
\begin{equation}
 \widehat F_\tau=\frac16\sum_{i=1}^3\sum_{w=0}^1
 \sum_{s:s_i=w}(p_s+p_{s^{(i)}})f_{i,w,s}^{(\tau)}.
 \label{eq:fluxfraction}
\end{equation}

Equation~\eqref{eq:fluxfraction} thus gives a single measure for comparing distributions: choose a face with equal probability, then a crossing according to its stationary weight; $\widehat F_\tau$ estimates the probability that the associated trajectory satisfies our instability criterion. The condition $s_i=w$ selects the outgoing configurations on that wall. Their weights sum to one on each face, and the factor $1/6$ averages the six results. This weighting accounts for how often each configuration crosses the wall, rather than treating all configuration labels as equally frequent. The resulting percentage measures the unstable fraction of this crossing ensemble, not the area occupied by orange points in one plot or a fraction of the cube's volume.

For each arrangement with three or four favored configurations, we test nine contrasts $r$ between 1 and 20, using 1536 initial conditions and testing growth at $T=3200$ and $12800$. At $r=5$, we increase the sample to 6144 initial conditions and test at $T=12800$ and $51200$. Samples comprise four independent scrambled Sobol replicates; reported uncertainties are the standard deviations of their flux estimates. Table~\ref{tab:classes} lists the resulting fractions. Its labels A1--A3 identify arrangements with three favored configurations, and B1--B6 identify those with four; the corresponding configurations are given explicitly in the second column. Within each group, the probability values are identical at fixed $r$, but assigning them to different basis states produces markedly different unstable fractions. All phases remain zero in this comparison.

\begin{table*}[t]
\caption{Geometric classes and checks at $r=5$, with 6144 initial conditions per class and $T=51200$. $\widehat F_{0.01}$ is the flux fraction exceeding the threshold at both $T=12800$ and $51200$; it is not an exact measure of asymptotic chaotic area. The sign $\pm$ denotes the standard deviation across four replicates. The final column gives the fraction with an intermediate diagnosis under the main criterion.}
\label{tab:classes}
\begin{ruledtabular}
\begin{tabular}{clrrr}
Class & Favored configurations & Multiplicity & $\widehat F_{0.01}$ & Unresolved\\
A1 & $000, 001, 010$ & 24 & $83.24\pm0.89\%$ & $6.04\%$\\
A2 & $000, 001, 110$ & 24 & $99.98\pm0.04\%$ & $0.02\%$\\
A3 & $000, 011, 101$ & 8 & $59.86\pm1.15\%$ & $2.50\%$\\
B1 & $000, 001, 010, 011$ & 6 & $0.00\pm0.00\%$ & $0.00\%$\\
B2 & $000, 001, 010, 100$ & 8 & $28.99\pm3.28\%$ & $18.10\%$\\
B3 & $000, 001, 010, 101$ & 24 & $99.78\pm0.08\%$ & $0.00\%$\\
B4 & $000, 001, 010, 111$ & 24 & $97.20\pm0.33\%$ & $0.03\%$\\
B5 & $000, 001, 110, 111$ & 6 & $0.00\pm0.00\%$ & $0.00\%$\\
B6 & $000, 011, 101, 110$ & 2 & $0.00\pm0.00\%$ & $0.00\%$\\
\end{tabular}
\end{ruledtabular}
\end{table*}

The opposite-vertex family $A=\{000,111\}$ also shows that stronger probability contrast need not increase instability. Sampling the flux on $y=0$ over its four outgoing labels, with 8192 initial conditions and horizons $T=12800,51200$, gives persistent fractions of $49.72\pm0.62\%$ at $r=2$ and $10.00\pm0.45\%$ at $r=20$. These single-face estimates, together with regular motion at $r=1$, establish a nonmonotonic response; they are distinct from the six-face averages in Table~\ref{tab:classes}.

The numerical distinction is less clear for some arrangements. In B2, the favored configurations are $000$ and its three nearest neighbors, $001$, $010$, and $100$; many trajectories have weak growth, making a longer observation important. In A1, with favored configurations $000$, $001$, and $010$, the estimated unstable fraction is sensitive to the chosen threshold. We therefore report these fractions as finite-time estimates, rather than exact chaotic areas. Even a fraction close to one can miss small regular islands. At $r=5$, raising the threshold from 0.005 to 0.02 lowers the persistent fraction in A1 from $86.95\%$ to $75.12\%$, whereas B2 changes from $29.85\%$ to $28.40\%$. For B2, the fraction exceeding 0.01 at a single horizon rises from $27.70\%$ at $T=3200$ to $30.16\%$ at $T=51200$.

The sections show where these different types of motion occur, complementing the percentages. The distributions A1 and A3 contain the same probability values, yet Fig.~\ref{fig:bornorganization}(c),(d) shows different positions of their regular islands and different patterns of unstable returns around them. For B2 [panel (e)], families of regular curves coexist with unstable returns; for B4 [panel (f)], unstable trajectories visit much of the displayed face, while small islands remain visible. Thus, rearranging the same Born probabilities changes both the frequency of instability and the geometry of the section. The figure illustrates this geometry on one face and configuration; the fractions in Table~\ref{tab:classes} use all six faces and all outgoing configurations.

These results also show that a three-body term in the probability model is neither sufficient nor necessary for instability. The regular family B6 favors the four configurations with even three-bit parity, $000$, $011$, $101$, and $110$. Its log probability contains a term proportional to $\sigma_x\sigma_y\sigma_z$, yet its motion is periodic. Conversely, B3 favors $000$, $001$, $010$, and $101$ and has
\begin{equation}
 \log p_{\rm B3}=\mathrm{const}
 +\frac{\log r}{4}(\sigma_x+\sigma_y+\sigma_x\sigma_z-\sigma_y\sigma_z),
\end{equation}

where $\sigma_i=1-2s_i$ and the constant fixes normalization. This expression contains only single-bit terms and two pairwise terms, with no three-body term. Nevertheless, at $r=5$, trajectories with persistent growth account for almost all of the sampled crossing weight. The relevant distinction is therefore not simply the highest interaction order: the particular relations among the eight probabilities determine whether the regularity conditions derived above hold.

\subsection{Eight distinct probabilities and decreasing profiles}
To check that the results do not rely on only two probability levels, we study profiles with eight distinct values. We label the computational-basis configurations by $k=4s_x+2s_y+s_z$, from 0 to 7, and set the probability contrast to $R=p_{\max}/p_{\min}$. A family of decreasing profiles is
\begin{equation}
 p_k\propto\exp[-(\log R)(k/7)^q],\qquad q>0.
 \label{eq:curvedprofile}
\end{equation}
The parameter $q$ controls the shape of the decrease while preserving the contrast $R$. For $q=1$, the exponential separates into three factors, $e^{-\beta k}=e^{-4\beta s_x}e^{-2\beta s_y}e^{-\beta s_z}$, with $\beta=\log R/7$. Each coordinate has period one. Thus, eight distinct decreasing probabilities can produce fully periodic motion.

\begin{figure*}[t]
\centering\includegraphics[width=.96\textwidth]{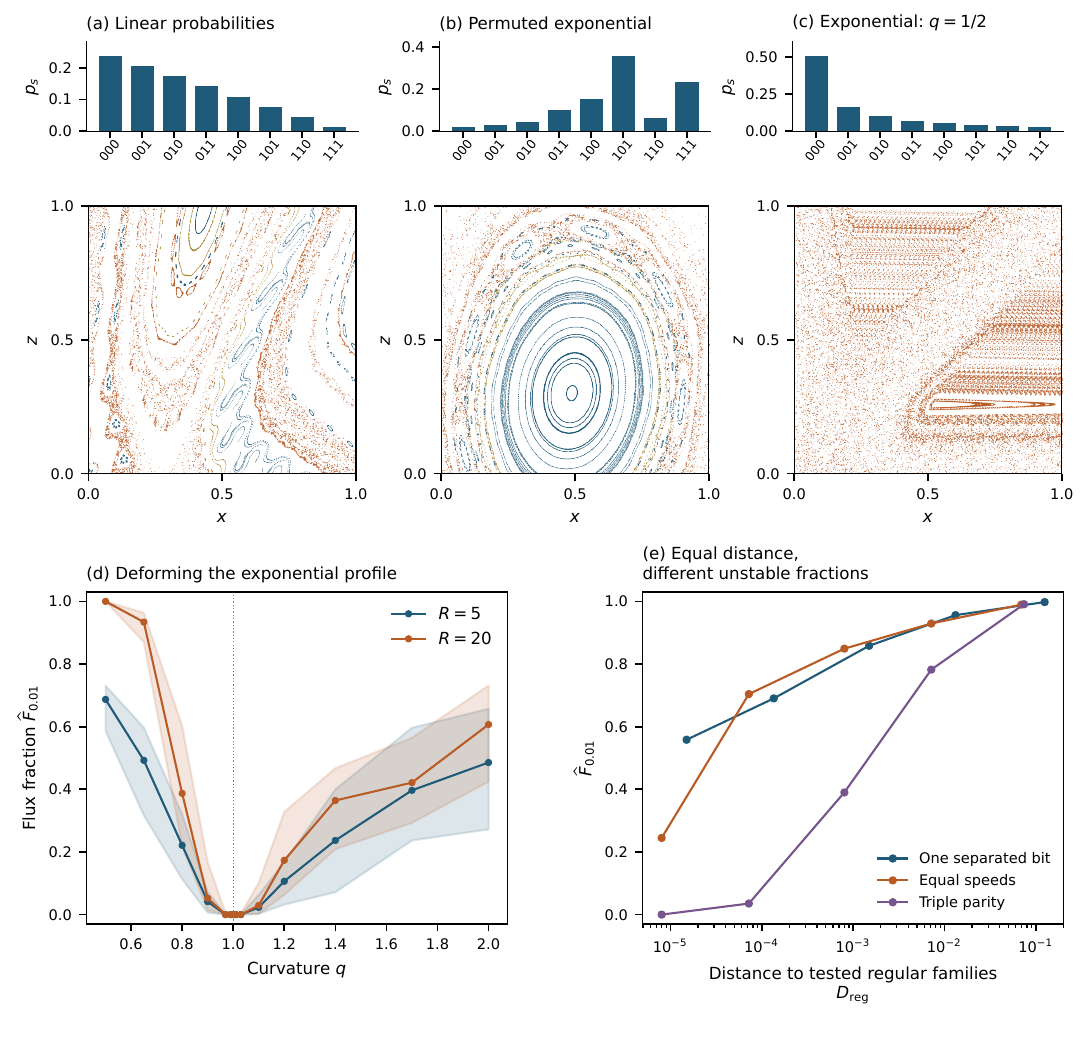}
\caption{Nontrivial structures with eight distinct probabilities. Panels (a)--(c) have contrast $R=20$. (a) Linearly decreasing probabilities. (b) Permutation of the exponential profile with $q=1$. (c) Deformed exponential profile with $q=1/2$. These three panels share the contrast but not the histogram: they compare island chains, a main island surrounded by unstable returns, and an unstable regime with return bands. All use 50 trajectories, $20<t\le800$, face $y=0$, and label $000$, with the colors of Fig.~\ref{fig:bornorganization}. (d) Deformation in Eq.~\eqref{eq:curvedprofile}. Bands vary the threshold from 0.005 to 0.02; they are not confidence intervals. (e) The same pairwise perturbation applied to three regular families, plotted against the minimum distance to the tested regular families. Curves (d),(e) use 1536 initial conditions per state up to $T=12800$ and average flux over six faces. Scalar proximity does not determine a unique unstable fraction.}
\label{fig:profiles}
\end{figure*}

Reassigning the same weights produces other dynamics; Fig.~\ref{fig:profiles}(b) shows islands and unstable trajectories for
\begin{equation}
 p_k\propto20^{-a_k/7},\quad (a_0,\ldots,a_7)=(7,6,5,3,2,0,4,1).
\end{equation} We enumerate the $8!=40320$ assignments of the exponential profile with $R=20$. Since all values are distinct, each assignment has 48 equivalent forms related by cube symmetries, leaving 840 distinct classes. Seven admit exact separation of one qubit. This property is checked using the integer exponents and holds for any finite $R$. Growth is detected in the other 833 classes in the exploratory sweep, with 192 initial conditions per class up to $T=3200$. The seven regular classes, the twelve with the smallest nonzero fraction, and twelve randomly selected classes are extended to $T=51200$ with 1536 initial conditions per class: 31 classes after removing duplicates. The enumeration is exhaustive; the instability diagnosis remains numerical.

Profiles linear in probabilities, linear in amplitudes, power-law profiles, and deformed exponentials complete a catalog of 132 cases. They exhibit regular regions, weak growth, and predominantly unstable regimes. Figure~\ref{fig:profiles}(a) shows linearly decreasing probabilities, $p_k\propto R-(R-1)k/7$: returns form island chains and unstable regions. The deformed exponential with $q=1/2$ [panel (c)] shows bands of unstable returns and a different spatial distribution. Unstable sections therefore do not have a unique geometry either. These differences in return density do not directly equal Born probabilities. Figure~\ref{fig:profiles}(d) shows reduced detected growth near $q=1$. Analytical factorization is established here only at $q=1$; a zero fraction at finite time does not prove integrability over an interval around that value.

\subsection{Interpreting instability in the representation}
There is no infinitesimal expansion during a flight because the velocity is constant. At an impact, two nearby trajectories reach the wall at slightly different times. If the configuration change also modifies velocities in other coordinates, this timing difference is transmitted between them. Its accumulation depends on the ordered product of impact matrices. The reductions above cancel expansion or limit it to shear; other sequences can accumulate exponential growth. A single distribution may support both sequences sustaining regular islands and unstable trajectories.

To quantify how much a distribution departs from the regular families identified above, we use the Kullback--Leibler divergence, $D_{\rm KL}(p\Vert q)=\sum_s p_s\log(p_s/q_s)$. It is zero when the two distributions coincide. We minimize it over two types of regular distributions: those in which one bit separates from the other two, $q=p_i p_{jk}$, and those unchanged by simultaneously flipping a pair of bits. For the first type, the closest distribution is the product of the corresponding marginals of $p$. For the second, it is $(p+p^{(ij)})/2$, where $p^{(ij)}$ is obtained by flipping bits $i$ and $j$ in every configuration label. We take the smallest value over all choices of the separated bit or flipped pair. We denote this minimum by $D_{\mathrm{reg}}(p)$, the horizontal coordinate in Fig.~\ref{fig:profiles}(e). This measures proximity to the tested regular families, but does not by itself predict the unstable fraction. For Fig.~\ref{fig:profiles}(e), we perturb three regular families using $p_s\propto\exp(\ell_s+\epsilon d_s)$, with $d_s=0.4\sigma_x\sigma_z-0.7\sigma_y\sigma_z$ and $\epsilon=0.01,0.03,0.1,0.3,1$. Their unperturbed log weights are
\begin{align}
 \ell_{\rm f}&=0.43\sigma_x-0.27\sigma_y+0.13\sigma_z
                 +0.8\sigma_x\sigma_y,\nonumber\\
 \ell_{\rm e}&=0.4\sigma_x+0.3\sigma_y\sigma_z
                 +0.6\sigma_x\sigma_y\sigma_z,\nonumber\\
 \ell_{\rm p}&=0.65\sigma_x\sigma_y\sigma_z.
 \label{eq:regularbases}
\end{align}
They describe, respectively, one separated bit, equal speeds, and triple parity. The distributions based on $\ell_{\rm e}$ and $\ell_{\rm p}$ at $\epsilon=0.03$ have the same minimum distance, $7.19948\times10^{-5}$, but give $\widehat F_{0.01}=72.99\pm1.30\%$ and $3.53\pm0.21\%$ with 6144 initial conditions up to $T=51200$. At threshold 0.002, they give $83.84\%$ and $18.48\%$. This contrast limits the interpretation: the specific probability relations and the starting point within a regular family matter, as well as the magnitude of their violation.
\section{Phase geometry}\label{sec:phasegeometry}
The angular extension allows states sharing a Born distribution to be compared. We now apply the interpolation rule to phase families and examine their geometry over the base trajectories. The relations identified here refer to this rule and to the combinations of coefficients that determine it.
\subsection{Angular structure for two qubits}
For two qubits, the phase interpolation is
\begin{align}
 U(x,y)={}&(1-x)(1-y)u_{00}+x(1-y)u_{10}\nonumber\\
 &+(1-x)y u_{01}+xyu_{11}.
 \label{eq:U2}
\end{align}
If phases are added independently to each qubit, $u_{s_1s_2}=e^{\iu(\alpha s_1+\beta s_2)}$, the field factorizes as $(1-x+xe^{\iu\alpha})(1-y+ye^{\iu\beta})$. For $|\alpha|,|\beta|<\pi$ there are no zeros, and the angle is the sum of a contribution from $x$ and one from $y$.

If only the phase of the coefficient of $|11\rangle$ is changed by $\gamma$, we obtain
\begin{equation}
 U=1+xy(e^{\iu\gamma}-1),\qquad
 \min_{[0,1]^2}|U|=|\cos(\gamma/2)|,
 \label{eq:controlled2}
\end{equation}
for $-\pi\le\gamma\le\pi$. In particular, $\gamma=\pi$ produces a curve of zeros, $xy=1/2$. Another choice, $(u_{00},u_{01},u_{10},u_{11})=(1,-\iu,\iu,-1)$, gives $U=1-x-y+\iu(x-y)$, which vanishes only at the center of the square. These choices produce different angular structures over the same Born dynamics. If a periodic orbit avoids zeros, its angle returns to the same value modulo $2\pi$ when the orbit closes, although it may accumulate one or more full turns along the path.

\subsection{Phase relations in three qubits}\label{sec:phases}
We now fix a three-qubit Born distribution and vary the phases of its coefficients. Spatial trajectories remain unchanged, allowing phase effects to be compared over the same base geometry. Any choice of three-qubit phases can be written, up to a global phase, as
\begin{equation}
 \phi_{\bs}=\sum_i\alpha_i s_i+
 \sum_{i<j}\beta_{ij}s_is_j+\gamma s_xs_ys_z
 \pmod{2\pi}.
 \label{eq:phaseexpansion}
\end{equation}
The coefficients $\alpha_i$ give the individual phase contributions, $\beta_{ij}$ describe contributions depending jointly on bits $i$ and $j$, and $\gamma$ gives the contribution involving all three bits. The expansion describes the eight state coefficients and is not an entanglement measure.

The probabilities and all Born trajectories remain fixed. The field $U$ in Eq.~\eqref{eq:U} changes the angular height of the same points. To isolate the joint phase dependence on $x$ and $z$, we compare the four coefficient phases on a face with fixed $s_y=y$, where $y=0$ or $1$. The combination
\begin{equation}
 \Gamma_{xz}(y)=\phi_{0y0}+\phi_{1y1}-\phi_{0y1}-\phi_{1y0}
 \pmod{2\pi}.
 \label{eq:facecontrast}
\end{equation}
cancels the global phase and all individual contributions $\alpha_i s_i$. Equivalently, it is the argument of $C_{0y0}C_{1y1}C_{0y1}^{*}C_{1y0}^{*}$. The coefficient magnitudes do not affect this argument. Substitution of Eq.~\eqref{eq:phaseexpansion} gives
\begin{equation}
 \Gamma_{xz}(0)=\beta_{xz},\qquad
 \Gamma_{xz}(1)-\Gamma_{xz}(0)=\gamma\pmod{2\pi}.
 \label{eq:phasecontrast}
\end{equation}
Thus, the face $y=0$ isolates the pairwise contribution $\beta_{xz}$, while comparing the two faces isolates its change $\gamma$ when the third bit switches from 0 to 1. Individual phases can still change the angular shape within a face; it is the vertex-phase combination $\Gamma_{xz}$ that remains unchanged by those contributions.

\subsection{A contribution invisible on one face and visible on the opposite face}
\begin{figure*}[t]
\centering\includegraphics[width=\textwidth]{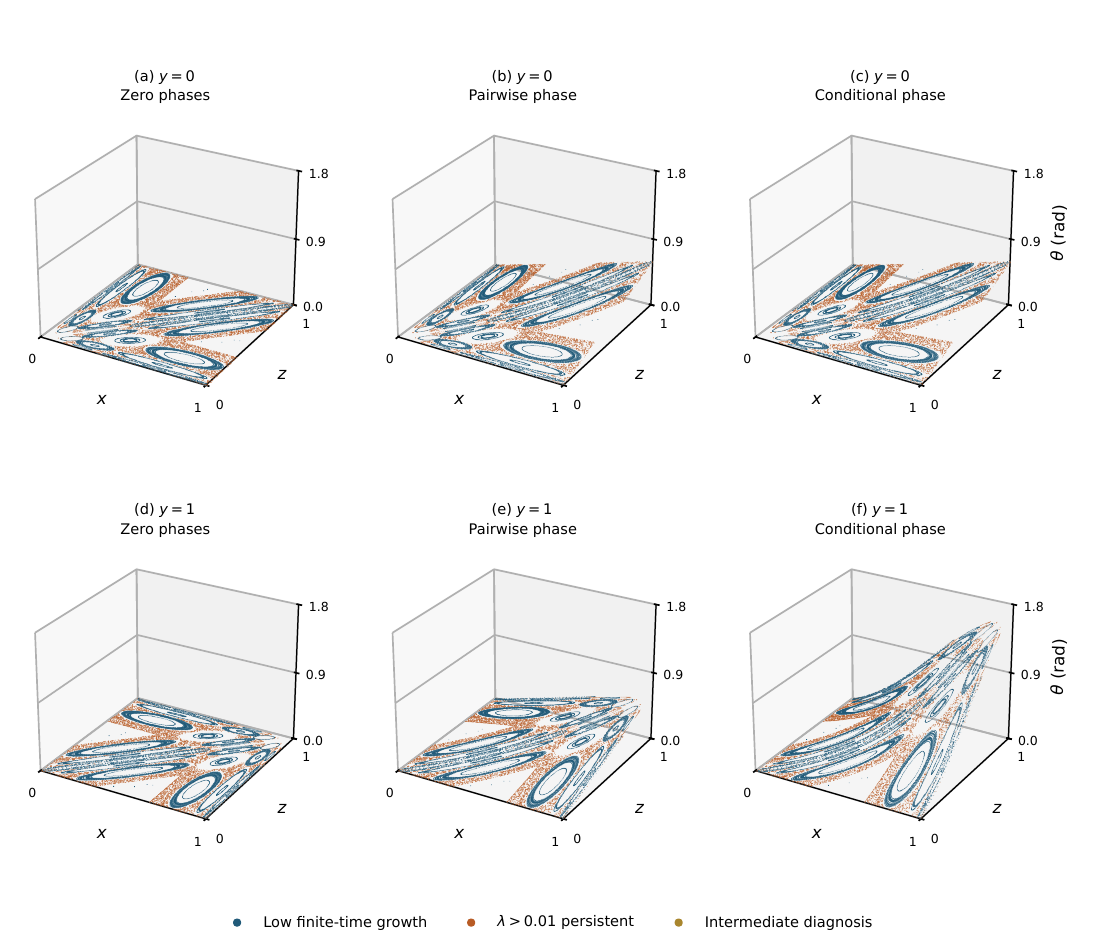}
\caption{Identical Born dynamics with different phase relations. The mixed Born case has $g=0.4$ in Eq.~\eqref{eq:symmetricborn}. Rows show $y=0$ with outgoing label $000$ and $y=1$ with label $010$. Columns correspond to $(\beta,\gamma)=(0,0),(0.6,0),(0.6,1.1)$. All columns use the same 50 original trajectories, four additional trajectories within regular islands, and their $x\leftrightarrow z$ images, giving a symmetric ensemble of 108 trajectories with matched return times and growth classification. Points with $20<t\le800$ are shown. The three-body phase leaves (b),(c) identical and changes the lift between (e),(f). Angular coordinates use the same scale in all panels. The additional and symmetry-related trajectories serve only to resolve the displayed structures; reconstruction tests use the original 50 trajectories.}
\label{fig:conditionalphase}
\end{figure*}

Consider the family $\phi_{\bs}=\beta s_xs_z+\gamma s_xs_ys_z$, where $\beta=\beta_{xz}$ and all other individual and pairwise phase terms vanish. Its interpolation can be written explicitly as
\begin{align}
 U(x,y,z)=1+xz\big[&(1-y)(e^{\iu\beta}-1)\nonumber\\
 &+y(e^{\iu(\beta+\gamma)}-1)\big].
 \label{eq:conditionalU}
\end{align}
On $y=0$, all dependence on $\gamma$ disappears. On $y=1$, the combination $\beta+\gamma$ determines the lift. Figure~\ref{fig:conditionalphase} compares zero phases, a pairwise phase $\beta=0.6$, and the addition of $\gamma=1.1$, over the mixed Born case of Eq.~\eqref{eq:symmetricborn} with $g=0.4$. Exactly the same trajectories are used in all three columns.

Panels (b) and (c) are identical: this face contains no information about the three-body phase. Panels (e) and (f) show how this contribution changes the angular height on the opposite face. Identical lifted sections do not imply identical states. Changing only $\gamma$ by $\Delta\gamma$ changes the phase of the coefficient of $|111\rangle$. The fidelity, defined as the squared overlap between the two normalized states, is
\begin{equation}
 |\langle\psi_\gamma|\psi_{\gamma+\Delta\gamma}\rangle|^2
 =1-4p_{111}(1-p_{111})\sin^2(\Delta\gamma/2).
 \label{eq:invisiblefidelity}
\end{equation}
Thus, distinct states can share the entire lifted section on $y=0$.

This comparison is repeated for an exponential product Born distribution with contrast $R=20$ and for the predominantly unstable B3 family at $r=5$. The visibility of $\gamma$ on the faces is unchanged because Eq.~\eqref{eq:conditionalU} depends on phases. Born probabilities change how trajectories explore that field. The construction therefore visualizes the same phase relations over different Born geometries. Independence of the base motion from phases is part of the CCBM rules.

In these tests, all vertex phase factors lie within an arc shorter than $\pi$ on the unit circle. Their interpolated weighted average $U$ therefore cannot vanish. The relation between coefficients and faces can be examined without singularities. The two-qubit examples above illustrate zeros and possible phase winding, whose geometry depends on the chosen interpolation. The connection used here is given by the phase combinations in Eq.~\eqref{eq:facecontrast}.
\subsection{Crossing a zero of the complex field}\label{sec:nodalcrossing}
A three-qubit product state allows a nodal crossing to be examined exactly:
\begin{equation}
 |\psi\rangle=|-\rangle\otimes|+\rangle\otimes|+\rangle,
 \qquad C_{\bs}=\frac{(-1)^{s_x}}{\sqrt{8}},
 \label{eq:nodalstate}
\end{equation}
where $|\pm\rangle=(|0\rangle\pm|1\rangle)/\sqrt{2}$. All probabilities are $p_{\bs}=1/8$. The phase factors are $u_{\bs}=(-1)^{s_x}$, and interpolation gives
\begin{equation}
 U(x,y,z)=1-2x,\qquad
 \zeta(t)=1-2x(t),\qquad \dot\zeta=-2v_x.
 \label{eq:nodaltrajectory}
\end{equation}
With $c_i=1$, all speeds are two. Each trajectory with isolated impacts repeatedly crosses the plane $x=1/2$, where $\zeta=0$. Its base motion is periodic and the complex lift remains continuous. The angle is zero for $x<1/2$ and $\pi$ modulo $2\pi$ for $x>1/2$; it is undefined on the nodal plane. The angular derivative vanishes on both sides and does not describe the change in argument across the node. Equation~\eqref{eq:nodaltrajectory} does describe the entire crossing.

This example distinguishes a singularity of the angular coordinate from trajectory instability. State information remains available in the velocities and vertex values of $U$. In the families shown in the main figures, the phase factors lie in an open semicircle, so the angular sections avoid this singularity. The crossing is specified completely by Eq.~\eqref{eq:nodaltrajectory}: flights change $\zeta$ linearly, and impacts preserve its value while changing its derivative.

\section{Reconstruction from machine records}\label{sec:reconstruction}
\subsection{Recovering phase relations from returns}\label{sec:recovery}
Reconstruction tests which information in the representation is retained by finite records. We recover phases from recorded positions and angles without supplying the fit with $\beta$ or $\gamma$. We fix $\phi_{000}=0$ and fit the other seven phases freely. Given the interpolation, we calculate $U$ at the observed positions and minimize the sum of squared real and imaginary differences between its angular factor $U/|U|$ and the measured $e^{\iu\theta}$. The procedure respects angular periodicity and the unit magnitudes of phase factors. We then calculate $\Gamma_{xz}(0)$, $\Gamma_{xz}(1)$, and their difference from the recovered phases.

The combinations $\Gamma$ refer to coefficient phases. In general, recovering them does not amount to directly subtracting the heights of two interior points on opposite faces; the known interpolation connects the records to their vertices. This fit tests the information preserved by the representation and is not tomography from quantum measurements.

We use the three Born cases above and 15 phase choices for each: $\beta=0.3,0.6,0.9$ and $\gamma=-0.9,-0.45,0,0.45,0.9$. We always add the terms $0.1s_x-0.08s_y+0.07s_z+0.12s_xs_y-0.09s_ys_z$ to check that individual and other pairwise contributions do not alter the identities in Eq.~\eqref{eq:phasecontrast}. The fit is not given this reduced structure; it still has seven unknowns.

Each test selects 8, 16, or 32 distinct trajectories from a collection of 50. From each trajectory, we take the first return after $t=20$ on each opposite face, including the different outgoing labels. Repetitions of a periodic orbit are not counted as new positions. Gaussian angular errors with standard deviations $0$, $10^{-3}$, or $10^{-2}$ radians are added, with eight realizations for each noisy case. There are 2295 fits in total, including 135 without noise.

To check how many phase parameters the records determine, we evaluate the Jacobian, the matrix of derivatives of the predicted angles with respect to the seven unknown phases. It has rank seven when both faces are combined: near the fitted solution, all seven phases can be distinguished. Using only $y=0$ gives rank three, because the four vertex phases with $s_y=1$ do not enter those records. This local test does not establish global uniqueness for every field or sample set.

Across the 135 noiseless tests, the maximum phase error, measured modulo $2\pi$, is below $1.3\times10^{-11}$ radians. We measure state error by the infidelity, one minus the fidelity. With 16 records per face and angular noise of $10^{-3}$ radians, median infidelities range from $5.3\times10^{-7}$ to $7.2\times10^{-7}$ across the three Born cases; for noise of $10^{-2}$ radians, they range from $4.7\times10^{-5}$ to $6.5\times10^{-5}$. These errors demonstrate accurate reconstruction for the families studied, both for regular motion and for cases with unstable regions. Precision depends on which positions are recorded and how strongly their angles constrain the fitted phases; chaos is not required for phase recovery.

Fits start from zero phases. As an additional control, 26 of 27 random initializations for three noiseless cases converge to the correct phases; one reaches a minimum with nonzero residual. The agreement between fitted and recorded angles, and the sensitivity of the fitted phases to errors, must therefore be checked; optimizer convergence alone does not guarantee reconstruction.

\subsection{Born and phase information in state reconstruction}
The reconstruction above keeps Born probabilities known to isolate phase effects. Probabilities can be obtained from occupation times under the conditions in Sec.~\ref{sec:occupation}, or from the velocity ratios developed in Appendix~\ref{app:inverse}. Once both parts are recovered, $C_{\bs}=\sqrt{p_{\bs}}e^{\iu\phi_{\bs}}$ gives the wave function.

As a complementary check of this full procedure, we test 64 states from a complex RBM with two hidden neurons, whose complex parameters have independent, zero-mean Gaussian real and imaginary parts, with standard deviation 0.25 for visible and hidden biases and 0.30 for couplings. For each state, we record 128 complete flights and their midpoint angles. Inversion of probabilities and phases from interior records gives maximum coefficient errors of $1.2\times10^{-14}$ and errors below $2.1\times10^{-14}$ over the 63 nonidentity Pauli expectation values without noise. Appendix~\ref{app:inverse} explains the sufficient rank-14 criterion for the linear problem in real and imaginary parts. Here the unknowns are the real and imaginary parts of seven phase factors, giving 14 real variables. The angular fit instead uses the seven phases directly; the two rank criteria refer to these different sets of unknowns.
\section{Discussion and conclusions}\label{sec:conclusions}
The complex chaotic Boltzmann machine extends Suzuki's construction to represent pure quantum states. Its complex parameters determine the wave-function coefficients; Born probabilities set the velocities, and relative phases define the complex field and its angular representation. The invariant distribution assigns each binary configuration its Born probability, while trajectories and their sections provide a geometry for exploring relations between coefficients.

The angular extension completes the original machine's probabilistic description. The chosen interpolation reproduces vertex phases, respects their periodicity and the cube symmetries, and factorizes when the phase factors do. These conditions provide a definite rule for lifting sections. State parameters therefore specify both the base motion and its angular representation without additional interpolation parameters.

The few-qubit analysis shows how to use this representation. Suzuki's reduction establishes the regular two-qubit reference. For three qubits, separating one qubit or imposing certain velocity equalities reduces the motion to rotations. These relations explain regular families that include correlated distributions. The calculations also reveal islands and unstable regions when the weight organization changes. The representation shows why keeping the same eight probability values does not preserve the dynamics when those values are assigned to different basis states.

Families with eight distinct probabilities extend this comparison. Exponential decay in binary order separates the three qubits and yields periodic motion, whereas other assignments of the same weights and other decreasing profiles produce mixed or predominantly unstable structures. In the sections, unstable trajectories may be distributed around islands, between families of curves, or in bands. Instability is interpreted through the transmission of arrival-time differences between coordinates at impacts. Distance to a regular family helps organize perturbations, but the response also depends on which relations are broken.

Phases add another dimension to this comparison. At fixed Born probabilities, their variation changes the lift of spatial structures. Restricting the field to opposite faces reveals how a pairwise phase contribution depends on the third qubit. Identities between coefficient phases make this geometric interpretation precise. Reconstruction tests with seven free phases and angular noise show that records from both faces can retain enough information to recover these relations. This check also succeeds for periodic Born dynamics: record coverage matters even when the base motion is regular.

The geometry must be interpreted within the representation's rules. We fix a computational basis, speed scales $c_i$, and an angular interpolation. Individual sections show the region explored by the selected trajectories; quantitative instability comparisons use six-face sampling and a finite-time criterion. Complete velocity and phase information determines the wave function, whereas finite records must sample enough distinct positions to determine the unknown coefficients. These distinctions allow the images and diagnostics to be used without treating them as a universal classification of states.

The construction is defined for any number of qubits with nonzero coefficients and trajectories with isolated impacts, although the detailed analysis is limited here to two and three qubits. The complex lift continues through field zeros; its angular representation is used away from them. Zeros caused by phase cancellation are distinct from zero wave-function amplitudes. The general parameterization and phase field contain exponentially many terms. The demonstrated utility is geometric; these results establish no compression or computational-cost advantage over other representations. Structured families may admit more compact expressions; extension to larger systems requires studying which parameterizations retain tractable evaluation of the dynamics and angular field. Zero amplitudes and mixed states require additional treatment. Neural density-matrix representations provide relevant groundwork for the latter problem~\cite{TorlaiMelko2018}.

The CCBM provides a way to parameterize and visualize quantum states through a deterministic system. Its contribution is the joint representation of Born probabilities and phases in a geometry of trajectories, together with an analysis showing how different coefficient families appear in that geometry. This perspective opens a route to studying state parameterizations through their dynamical structures.

\appendix
\section{Connection to restricted Boltzmann machines}\label{app:rbm}
RBMs provide an alternative coefficient parameterization that can be used to evaluate the CCBM. NQS have been applied to time evolution and state reconstruction from measurements~\cite{Schmitt2020,Torlai2018,Carrasquilla2019,Neugebauer2020}. Here we use this connection to show how velocities can be evaluated from a network and to test reconstruction with states generated by it.

A restricted Boltzmann machine has a layer of visible variables, associated here with the qubits, and a layer of hidden variables. Connections join the two layers, with no couplings within either layer. Summing over the hidden variables produces an amplitude for each visible configuration. With $\sigma_i=1-2s_i$, a complex RBM assigns configuration $\bm\sigma$ the amplitude
\begin{align}
 \Psi(\bm\sigma)&=\sum_{\bm\eta}\exp\!\left(
 \bm a\cdot\bm\sigma+\bm b\cdot\bm\eta+
 \sum_{ji}\eta_jW_{ji}\sigma_i\right)\nonumber\\
 &=e^{\bm a\cdot\bm\sigma}\prod_j
 2\cosh\!\left(b_j+\sum_iW_{ji}\sigma_i\right).
 \label{eq:rbm}
\end{align}
The hidden variables $\eta_j$ take values $\pm1$ and are summed in the first line. The second line performs this sum explicitly. Biases $a_i,b_j$ and weights $W_{ji}$ may be complex. The visible biases of this RBM and the individual coefficients in the expansion of $L$ belong to different parameterizations and generally do not coincide. The normalized amplitude is $C_{\bs}=\Psi(\bm\sigma)/\sqrt{\sum_{\bm\sigma'}|\Psi(\bm\sigma')|^2}$, with the global phase fixed by the reference.

For the positive family in Eq.~\eqref{eq:symmetricborn}, with $g\ge0$, one hidden variable suffices to write
\begin{align}
 \Psi_g(\bm\sigma)&\propto
 2\cosh[W(\sigma_x+\sigma_y+\sigma_z)],\nonumber\\
 W&=\operatorname{arcosh}\sqrt{\frac{e^{2g}+3}{4}}.
 \label{eq:familyrbm}
\end{align}
For the two states with equal signs the sum is $\pm3$; for the other six it is $\pm1$. The identity $\cosh(3W)/\cosh W=e^{2g}$ gives precisely their amplitude ratio. Phases added later are not restricted to this same minimal architecture.

For the sizes studied here, the visible expansion in Eq.~\eqref{eq:loggeneral} ends at
\begin{align}
 L_2&=a_1s_1+a_2s_2+J_{12}s_1s_2,\nonumber\\
 L_3&=\sum_i a_i s_i+\sum_{i<j}J_{ij}s_i s_j+K s_1s_2s_3.
 \label{eq:visible}
\end{align}
The coefficients are obtained by finite differences of $L$ over binary configurations. In particular,
\begin{align}
 K={}&L(111)-L(110)-L(101)-L(011)\nonumber\\
 &+L(100)+L(010)+L(001)-L(000).
 \label{eq:K}
\end{align}
This term is necessary for a general three-qubit parameterization. In the symmetric Born family of Eq.~\eqref{eq:symmetricborn}, the log magnitudes contain only individual and pairwise terms, so their contribution to $K$ vanishes. Other families may have a real third-order contribution. The phases in Eq.~\eqref{eq:phaseexpansion} contribute $\operatorname{Im}K=\gamma$ for the chosen representatives. Both Born and phase contributions are thus included in the general CCBM parameterization.

In the RBM of Eq.~\eqref{eq:rbm}, $W_{ji}$ couples visible and hidden units. Effective visible interactions, including $K$ when present, are obtained after summing over hidden variables and taking the logarithm. The expansion coefficients depend on the branches chosen for the phases, whereas the state is fixed by $e^{L(\bs)}$.

\section{Reconstruction from finite records}\label{app:inverse}
\subsection{Probabilities from velocities}
Equation~\eqref{eq:flow} can be solved for a probability ratio:
\begin{equation}
 \frac{p_{\bs^{(i)}}}{p_{\bs}}=
 \frac{|v_i(\bs)|}{c_i}-1.
 \label{eq:inverse}
\end{equation}
The absolute value removes the velocity sign, and division by $c_i$ removes its scale. The constant term is subtracted when solving Eq.~\eqref{eq:flow} for the ratio. The resulting expression gives a probability ratio between neighboring configurations without a time average.

Reconstruction assigns a provisional weight to $\bm0$ and obtains the others by multiplying ratios along single-bit flips. Normalization by the sum of weights determines the probabilities. It is sufficient for the observed relations to connect all configurations; agreement along different paths checks consistency.

For a flight record, each velocity is obtained from the difference between final and initial positions divided by the duration. When a label is observed more than once, our tests use its longest flight to reduce rounding effects. To evaluate the dynamics directly from an RBM, the amplitude ratio for a sign flip is
\begin{align}
 \frac{\Psi(\bm\sigma^{(i)})}{\Psi(\bm\sigma)}=
 e^{-2a_i\sigma_i}\prod_j
 \frac{\cosh(\zeta_j-2W_{ji}\sigma_i)}{\cosh\zeta_j},\nonumber\\
 \zeta_j=b_j+\sum_kW_{jk}\sigma_k.
 \label{eq:rbmratio}
\end{align}
Its squared magnitude gives the Born ratio and avoids calculating the global normalization when evaluating velocities.

\subsection{Phases from positions and angles}
If the full angular field is known, its vertex values directly give $u_{\bs}=e^{\iu\phi_{\bs}}$. Together with the probabilities above, these recover the wave function up to a global phase. Interior zeros do not prevent reconstruction because the field has unit magnitude at the vertices.

Finite records, however, usually contain angles at interior positions. Suppose a position $\bx_k$ and its angle $\theta_k$ are known. Rotating $U(\bx_k)$ by $-\theta_k$ must give a positive real number, so its imaginary part must vanish. Substituting the interpolation and fixing $u_0=1$ gives
\begin{align}
 &\sum_{s\ne0}B_s(\bx_k)
 [-\sin\theta_k\operatorname{Re}u_s
 +\cos\theta_k\operatorname{Im}u_s]\nonumber\\
 &\hspace{20mm}=B_0(\bx_k)\sin\theta_k.
 \label{eq:linear_inverse}
\end{align}
Here $s$ enumerates binary configurations, and $B_s(\bx_k)$ is the interpolation weight defined in Eq.~\eqref{eq:U}. Each measured position and angle supplies one linear equation for the phase factors. Once the reference factor is fixed, there are $2(2^n-1)$ real unknowns: the real and imaginary parts of each phase factor. If the sampling matrix has full column rank, the solution is unique. For three qubits this rank is 14. The exact solution must have unit magnitude for each factor and also satisfy
\begin{equation}
 \operatorname{Re}\!\left[e^{-\iu\theta_k}
 \sum_sB_s(\bx_k)u_s\right]>0.
 \label{eq:oriented}
\end{equation}
This inequality checks that the field points in the measured direction rather than the opposite one. The original state satisfies all equations; if the linear solution is unique, it coincides with that state. This is a sufficient condition. Samples restricted to one face, or particular phase choices, may not yield this rank.

In the 64 tests, the matrix has rank 14 with a threshold of $10^{-11}$ times its largest singular value. Its largest condition numbers are $463.7$ for flight samples and $536.7$ for volume samples. This quantity indicates how much errors can be amplified when solving the system. With noise, we use least squares and divide each recovered factor by its magnitude. Finally, we form $C_{\bs}=\sqrt{p_{\bs}}u_{\bs}$. These tests reconstruct states from machine records; neural tomography addresses the different problem of inferring states from measurement outcomes~\cite{Torlai2018,Carrasquilla2019,Neugebauer2020}.

\section{Basis changes and expectation values from the representation}\label{sec:observables}
Born probabilities and relative phases determine the coefficients $C_{\bs}=\sqrt{p_{\bs}}e^{\iu\phi_{\bs}}$. Phases can be obtained from the vertices of the complex field or from sufficient records, as developed in Appendix~\ref{app:inverse}. Having both parts allows the same state to be expressed in any orthonormal basis and its probabilities to be calculated there.

Let $\{|b_\alpha\rangle\}$ be a new basis, with $\alpha=0,\ldots,2^n-1$, and let $M_{\alpha\bs}=\langle b_\alpha|\bs\rangle$ be the unitary change-of-basis matrix. The corresponding amplitudes and probabilities are
\begin{align}
 D_\alpha&=\langle b_\alpha|\psi\rangle
 =\sum_{\bs}M_{\alpha\bs}\sqrt{p_{\bs}}e^{\iu\phi_{\bs}},\nonumber\\
 q_\alpha&=|D_\alpha|^2.
 \label{eq:changebasis}
\end{align}
The procedure is to recover the complex coefficients, multiply them by $M$, and take the squared magnitude of each component. Phases enter the amplitude sum and determine interference. Two states with the same Born distribution in the computational basis can therefore have different probabilities in another basis, even when they share the CCBM base dynamics.

For a Hermitian observable $O$, we may choose $|b_\alpha\rangle$ as its eigenvectors and denote its eigenvalues by $o_\alpha$. Its expectation value is then
\begin{equation}
 \langle O\rangle=\sum_\alpha o_\alpha q_\alpha.
 \label{eq:spectralmean}
\end{equation}
The basis change uses the recovered state information and requires no further trajectory propagation. To construct a CCBM in that new basis, its probabilities $q_\alpha$ would determine different velocities through Eq.~\eqref{eq:flow}, provided that all coefficients are nonzero. The quantum basis change remains well defined when a transformed amplitude vanishes, although the dynamical construction studied here excludes that case.

When $O$ is already diagonal in the computational basis, with value $f(\bs)$ for each label, it suffices to weight the flight times:
\begin{equation}
 \widehat{\langle f\rangle}=
 \frac{\sum_{a,k}\tau_{ak}f(\bs_{ak})}
 {\sum_{a,k}\tau_{ak}}.
 \label{eq:timeobs}
\end{equation}
Here $a$ identifies the trajectory, $k$ one of its flights, $\tau_{ak}$ its duration, and $\bs_{ak}$ its label. The estimator requires the same sampling conditions as Born estimation. For a general operator, diagonalization can be avoided by using the local estimator~\cite{Medvidovic2024},
\begin{align}
 O_{\rm loc}(\bs)&=\sum_{\br}O_{\bs\br}
 \frac{C_{\br}}{C_{\bs}},\nonumber\\
 \langle O\rangle&=\sum_{\bs}p_{\bs}O_{\rm loc}(\bs),
 \label{eq:observable}
\end{align}
where $O_{\bs\br}=\langle\bs|O|\br\rangle$. In this case, $O_{\rm loc}$ is calculated for each label from the recovered amplitudes and summed with Born weights, or used as the weight of each flight under the same sampling conditions.

For example, $X_i$ flips bit $i$ and has eigenvalues $\pm1$. The real part of its local estimator is
\begin{equation}
 f_{X_i}(\bs)=
 \sqrt{\frac{p_{\bs^{(i)}}}{p_{\bs}}}
 \cos(\phi_{\bs^{(i)}}-\phi_{\bs}).
 \label{eq:xlocal}
\end{equation}
Its mean gives $\langle X_i\rangle$, and the probabilities of the two outcomes are $(1\pm\langle X_i\rangle)/2$. This expression shows how phase differences change measurement probabilities in another basis without changing the flight times of the original dynamics. Reconstruction checks use the exact sum over configurations, separating reconstruction error from sampling error.

\section{Perturbation growth at impacts}\label{app:stability}
Two nearby trajectories have the same velocity between impacts if they carry the same label. Their separation remains unchanged during that flight. However, they reach the face at slightly different times. During this time difference, one already has the post-impact velocity while the other still has the pre-impact velocity. This contribution must be included when calculating perturbation dynamics.

For an impact in coordinate $i$, the matrix relating perturbations at a common time before and after impact is~\cite{Kong2024}
\begin{equation}
 S=I+\frac{(\bv^+-\bv^-)\bm e_i^{\mathsf T}}{v_i^-}.
 \label{eq:saltation}
\end{equation}
Here $I$ is the identity, $\bv^-$ and $\bv^+$ are the pre- and post-impact velocities, and $\bm e_i$ is the unit normal to the face. We propagate a small perturbation by multiplying it by $S$ at each impact and renormalize it to avoid overflow. The sum of the logarithms of its growth factors, divided by the observation time $T$, gives $\lambda_T$. In our tests, this derivative agrees with finite differences at short times to within $1.2\times10^{-9}$.

\bibliography{references}
\end{document}